# Modeling of microfocused Brillouin light scattering spectra

Ondřej Wojewoda[,*] Martin Hrtoň, and Michal Urbánek[,†]

*CEITEC BUT, Brno University of Technology, Purkyňova 123, Brno 612 00, Czech Republic*

Although micro-focused Brillouin light scattering (BLS) has been used for more than 20 years, it lacks a complete theoretical description. This complicates the analysis of experimental data and significantly limits the information that can be obtained. To fill this knowledge gap, we have developed a semi-analytical model based on the mesoscopic continuous medium approach. The model consists of the following steps: calculation of the incident electric field and the dynamic susceptibility, calculation of the induced polarization, and calculation of the emitted electric field and its propagation towards the detector. We demonstrate the model on the examples of the measurements of thermal and coherently excited spin waves. However, the used approach is general and can describe any micro-focused Brillouin light scattering experiment. The model can also bring new analytical approaches to mechanobiology experiments or to characterization of acoustic wave-based devices.

## I. INTRODUCTION

Brillouin light scattering (BLS) is the inelastic scattering of a photon on a quasiparticle, usually a phonon or magnon [1,2]. However, due to the small frequency shift (typically in the range of several hundred megahertz to tens of gigahertz), it was a major challenge to confirm it experimentally. With the advent of the tandem Fabry-Perot interferometer [3,4], which allowed the measurement of such small frequency shifts with very high contrast ($10^{15}$), the BLS became the main tool for the study of phonons (acoustic waves) and magnons (spin waves). The typical BLS experiment was performed on bulk samples or thin films. In order to optimize the signal, the measurement was usually performed with the laser beam spanning tens of micrometers. Such a large spot can be regarded as a homogeneous plane wave, and thus the theory developed to describe the experimental data used this as an integral assumption [5–7].

However, at the beginning of the millennium, the advent of nanofabrication techniques sparked interest in diffraction-limited spatial scanning of quasiparticles. This was addressed by so-called micro-focused Brillouin light scattering ($\mu$-BLS) [1,2]. In this imaging microscopy technique, the long working distance lens has been replaced by an objective lens with a high numerical aperture. Such small probing spots, combined with an automated 3D scanning stage, allowed rapid progress in the field of spin wave research by studying spin waves in nanostructures [8–12], spatially mapping their propagation characteristics, or even making time-resolved measurements. The same methodology has also been applied to the study of acoustic waves [13]. The spatial study of spontaneously excited acoustic waves is also of great interest in, e.g., mechanobiology, cancer treatment, or pharmaceutical research [14–17]. The analysis of micro-BLS data is usually based on fitting the position and width of the Brillouin peak. However, due to the narrow focus of the light provided by the high numerical aperture of the objective lens, the homogeneous plane wave approach cannot be used as in the case of conventional BLS because it does not take into account the spectral broadening and also does not correctly predict the peak shift. Antonacci *et al.* used weighted incoherent superposition of plane waves to account for spectral broadening of the BLS peak [18]. However, this approach is suited only for simple systems with a linear dispersion and since it disregards the vector nature of electromagnetic fields and magnetization, it fails to capture effects associated with polarization, e.g., it does not provide polarization of the scattered wave. To correctly model and analyze BLS data of magnetic layers with complex dispersion relations or, in general, any sample with nontrivial light-matter coupling mechanisms, a more thorough approach is required (see Fig. 1).

Here we address the lack of appropriate theoretical description by presenting a semi-analytical model for the calculation of the resulting signal in $\mu$-BLS experiments. Our approach is based on the mesoscopic description of inelastic scattering with a small frequency shift described by Landau and Lifschitz [19]. The modeling follows the approach used for the model of Mie-enhanced BLS [20,21]. However here, in the case of a bare magnetic thin film, we formulate everything using the semi-analytical approach and thus the $\mu$-BLS signal can be calculated within seconds which means that the model can be used for fitting and optimization of the studied systems. To model the $\mu$-BLS spectra we use the approach presented in [19, p. 413] applied to microfocused light. The model can be divided to four steps, see flow chart in Fig. 1.

---

*Contact author: ondrej.wojewoda@vutbr.cz
†Contact author: michal.urbanek@ceitec.vutbr.cz

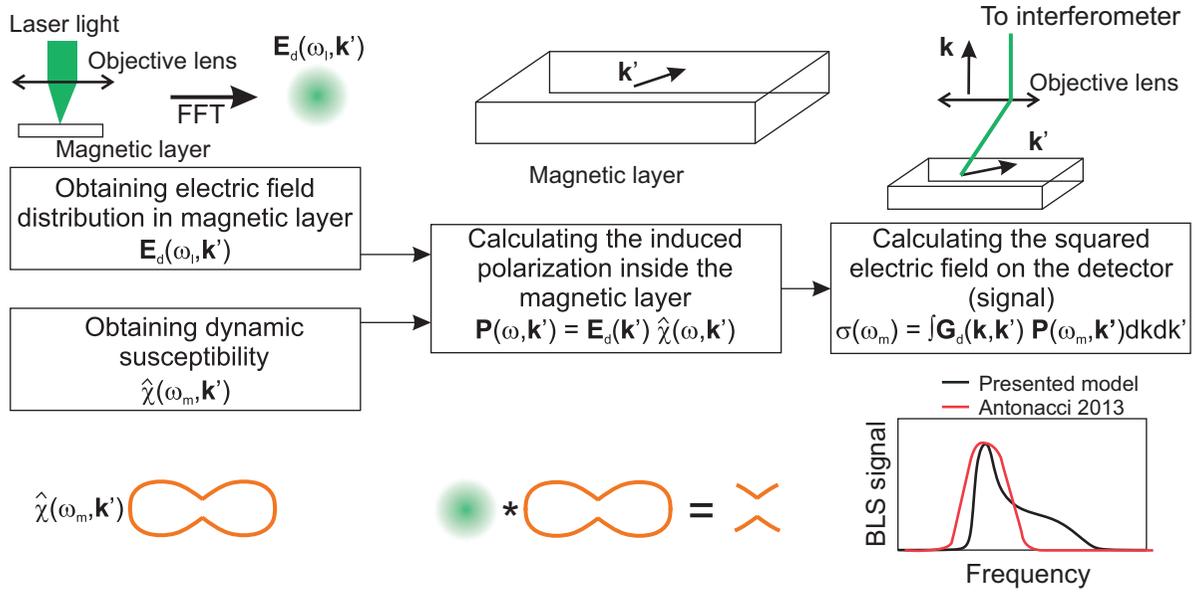

FIG. 1. Schematics of the presented model and comparison of its output with simplified approach using weighted superposition of plane waves. The **k'** represents the wavevectors in the plane of a magnetic layer, while **k** represents the wavevectors of scattered free light. The schematics provides a simplified picture and is not fully complete. For a more detailed discussion, please follow the explanation given in the text.

The incident electric light is calculated for an ideal objective lens using the theory developed by Richards and Wolf [22,23]. The dynamic susceptibility is calculated by assuming the Lorentzian form of the spin-wave resonance and using the zeroth perturbation theory of Slavin and Kalinikos [24,25], but in principle any formulation of the dispersion relation can be used, such as semi-analytical approaches [26] or simulation [27–29]. The incident electric field and the dynamic susceptibility are then combined [via convolution (multiplication) for calculations in reciprocal (real) space] to give the polarization induced within the magnetic layer. Finally, the angular spectrum of the radiation emitted by the polarization current and collected by the objective is calculated using the Green's function formalism developed by Sommerfield [30] and extended by Weyl [31].

For simplicity, in the following text we demonstrate our model of $\mu$-BLS on spin waves in a single 30 nm-thick (unless otherwise stated) Permalloy layer. However, any configuration of top/bottom layers can be assumed.

The paper is structured as follows: Section II presents the overall structure of the model. Section III shows in detail the procedure to obtain the incident electric field. Section IV shows the calculation of the dynamic susceptibility. The induced polarization is discussed in Sec. V. In Sec. VI the electric field at the detector is obtained and the formation of the detected signal is discussed. Section VII shows modeled and experimentally obtained spectra of thermally and coherently excited spin waves.

## II. STRUCTURE OF THE MODEL

### A. Calculation of the electric field distribution in the magnetic layer

The first step is to calculate the electric field in the sample ($E_d$). Usually this is easier to do in real space coordinates.

However, for subsequent calculations it is much more convenient to transform it into Fourier space, where we can write

$$\boldsymbol{E}_d(\omega_l, \boldsymbol{k}') = \mathcal{F}(\boldsymbol{E}_d(\omega_l, \boldsymbol{r}')), \quad (1)$$

where $\omega_l$ is a frequency of light. The shape of the electric field depends, for example, on the lens used (especially its numerical aperture), the wavelength of the light used, or the defocus of the light. In addition, the substrate, the magnetic layer material or some capping layers (e.g., antireflection coatings) can modify the electric field distribution, but in the presented approach we neglect these contributions. If necessary, a numerical calculation of Maxwell equations using finite-elements or finite differences methods can take these effects into account.

### B. Calculation of the dynamic susceptibility

The dynamic susceptibility ($\hat{\boldsymbol{\chi}}$) is an integral part of the model for calculating BLS spectra. This quantity represents the mechanism of interaction between the light and the excitation of the matter. In the specific case of the interaction between the spin waves and the light, this interaction is called magneto-optical coupling. To calculate the dynamic susceptibility, one needs to know the magnetization distribution as a function of the wavevector and the frequency of the spin waves

$$\hat{\boldsymbol{\chi}} \propto \boldsymbol{M}(\omega_m, \boldsymbol{k}'). \quad (2)$$

In the approach presented here, we use a simplified model of Lorentzian oscillators in each $\boldsymbol{k}'$ and linear spin-wave dispersion. However, if necessary, the full micromagnetic simulation can be used to obtain the BLS spectra in (almost) any scenario.

### C. Calculation of polarization

The driving field and the dynamic susceptibility are used to calculate the induced polarization in the magnetic layer by their convolution.

$$\mathbf{P}(\omega, \mathbf{k}) = \mathbf{E}_d(\omega, \mathbf{k}) * \hat{\chi}(\omega, \mathbf{k}). \quad (3)$$

In general, the polarization can be induced with wavevectors double the incident driving field, see the convolution in Eq. (3). However, the possible states of the polarization in $k$-space are still limited by the eigenstates of the spin waves, i.e., they can only occur at the positions of the spin wave resonance.

### D. Transfer to the far field and calculation of the BLS signal

Now we need to propagate the polarization into the far field. To do this, we use the Green's function formalism. In our case, the polarization source can be considered as the sum of the dipoles (point sources); thus, by integrating over them, the total electric field can be obtained. This can be expressed as

$$\mathbf{E}_{\text{FF}}(\omega, \mathbf{k}_p) = \hat{\mathbf{G}}(\omega, \mathbf{k}_p, \mathbf{k}'_p)\mathbf{P}(\omega, \mathbf{k}'_p). \quad (4)$$

However, this equation describes the emission of all wavevectors, including those that do not reach the detector. For this reason, subsequent trimming of the electric field is required, e.g., trimming by the numerical aperture of the objective lens, or assuming only wavevectors parallel to the optical axis after passing through the objective lens.

To obtain the resulting BLS signal, the electric field must be squared. The resulting quantity can then be fitted to or compared with the acquired BLS signal.

## III. CALCULATION OF THE ELECTRIC FIELD DISTRIBUTION IN THE MAGNETIC LAYER

In this section we use the semi-analytical method developed by Richards and Wolf [22,23,32] to obtain the electric field on the surface. We assume an ideal lens and a collimated beam. For the calculation we use spherical coordinates defined by angles $\phi$ and $\theta$, but later we transform the solution to Cartesian coordinates for more convenient calculation, see Fig. 2.

The field is given by the equation

$$\mathbf{E}(\rho, \varphi, z) = \frac{ik_0 f^2}{2}\sqrt{\frac{n_1}{n_2}} E_0 e^{(-ik_0 f)} \begin{pmatrix} I_{00} + I_{02}\cos 2\varphi \\ I_{02}\sin 2\varphi \\ -2iI_{01}\cos\varphi \end{pmatrix}, \quad (5)$$

where $k_0 = \frac{2\pi}{\lambda}$ is the wave number of the laser light, $\lambda$ is its wavelength, $n_1$ ($n_2$) is the index of refraction of the surrounding medium (sample under investigation), $E_0$ is incident electric field intensity, $f$ is an effective focal distance [33]. The integrals ($I_{00}$, $I_{01}$, and $I_{02}$) are given by

$$I_{00} = \int_0^{\Theta_{\text{max}}} f_w(\Theta)(\cos\Theta)^{1/2}\sin\Theta(1+\cos\Theta) \\ \times J_0(k\rho\sin\Theta)\exp(ik_0 z\cos\Theta)d\Theta, \quad (6a)$$

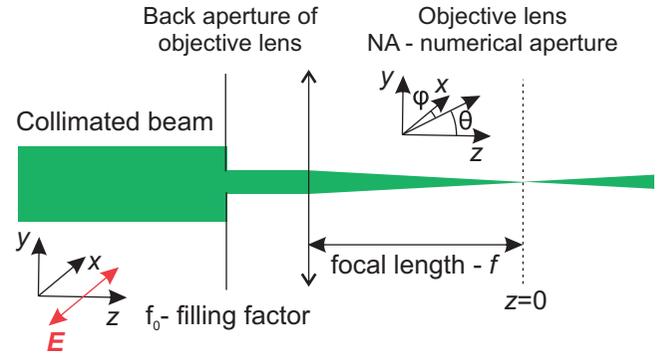

FIG. 2. Sketch of the calculation geometry for focal fields. In the region of the collimated beam we use Cartesian coordinates as shown in the sketch. The filling factor is the ratio between the sizes of the incident beam and the back aperture (entrance pupil) of an objective lens. The beam profile can be calculated in arbitrary $z$-plane.

$$I_{01} = \int_0^{\Theta_{\text{max}}} f_w(\Theta)(\cos\Theta)^{1/2}\sin^2\Theta \\ \times J_1(k\rho\sin\Theta)\exp(ik_0 z\cos\Theta)d\Theta, \quad (6b)$$

$$I_{02} = \int_0^{\Theta_{\text{max}}} f_w(\Theta)(\cos\Theta)^{1/2}\sin\Theta(1-\cos\Theta) \\ \times J_2(k\rho\sin\Theta)\exp(ik_0 z\cos\Theta)d\Theta, \quad (6c)$$

where $J_n(x)$ is the *Bessel* function of the $n$th-order $f_w(\Theta) = \exp(\frac{-1}{f_0^2}\frac{\sin^2\Theta}{\sin^2\Theta_{\text{max}}})$, $f_0$ is a filling factor (ratio between the sizes of the incident beam and the back aperture of the objective lens), $\rho$ is a distance from center in polar coordinates. The field is then transformed to the Cartesian coordinates.

The results for a laser wavelength of 532 nm, an objective lens with a numerical aperture of 0.75 and a fill factor of 2 are shown in Figs. 3(a)–3(c). The magnitude of the $x$ component [Fig. 3(a)] has full polar symmetry, while the magnitudes of the $y$ component [Fig. 3(b)] and the $z$ component [Fig. 3(c)] resemble fourfold and twofold symmetries respectively. The magnitude of the $x$ component is the largest. The maximum value of the square field in the $z$ component is ≈12.5× and the $y$ component is ≈715× smaller than the $x$ component. Thus, the $x$ component (along the polarization axis) is the most important for the calculation of the BLS signal in most scenarios. However, in certain geometries (e.g., for the out-of-plane magnetized thin film), its contribution may cancel out and other components may become significant.

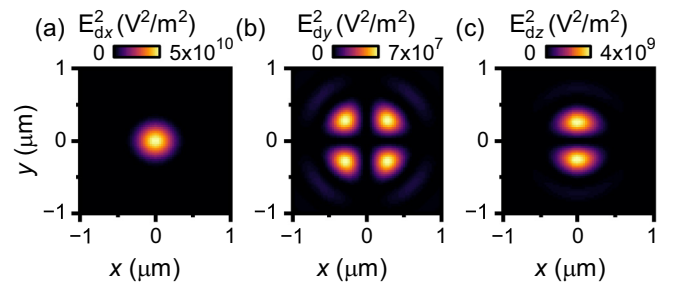

FIG. 3. Semi-analytically calculated magnitude of Gaussian beam. (a)–(c) The squared $x$- (a) $y$- (b) and $z$- (c) component of the focal field of 00 Hermite-Gaussian mode.

## IV. CALCULATION OF THE DYNAMIC SUSCEPTIBILITY

This section introduces the procedure for obtaining *Bloch spectral density of states* of thermally excited spin waves. The term *Bloch spectral density of states* or *Bloch function* [$\mathcal{D}(\omega, \boldsymbol{k})$] is often used for the function describing the density of states in frequency-wavevector space, i.e., [34,35]. We present here a phenomenological approach, which can be used together with any method for obtaining dispersion relation together with the calculated (or estimated) lifetime of the spin waves. This allows one to choose an appropriate method for obtaining the dispersion relation, including numerical calculation, since the fully analytical method may give incorrect results for, e.g., thick magnetic layer in the dipolar-exchange region (where the product of the wave vector and the sample thickness is comparable to 1).

To calculate the dispersion relation we use the zeroth perturbation model [24,25]:

$$\omega^2 = (\omega_H + A^2 \omega_M k^2)(\omega_H + A^2 \omega_M k^2 + \omega_M F_n), \quad (7)$$

where $\omega = 2\pi f$ is the spin-wave frequency, $k$ is its wavevector, $\omega_H = \mu_0 \gamma H_{ext}$, $\omega_M = \mu_0 \gamma M_s$, $M_s$ is the saturation magnetization, $\gamma$ is the gyromagnetic ratio, $\mu_0$ is the permeability of a vacuum, $A$ is the exchange length,

$$F_n = P_n + \sin(\theta)^2 (1 - P_n[1 + \cos(\phi)^2]) + \frac{\omega_M P_n (1 - P_n) \sin(\phi)^2}{\omega_H + l_{ex}^2 \omega_M k^2}, \quad (8)$$

where $\varphi$ is the in-plane angle, $\vartheta$ is the out-of-plane angle,

$$P_n = \begin{cases} \frac{Q^2}{k^2} - \frac{Q^4}{k^4} \frac{1}{2} \left[ \frac{2}{Qt}(1 - \exp(-Qt)) \right] & n = 0 \\ \frac{Q^2}{k^2} - \frac{Q^4}{k^4} \left[ \frac{2}{Qt}(1 - \exp(-Qt)) \right] & n \neq 0 \end{cases}, \quad (9)$$

and $t$, is the thickness of the sample. The spin-wave lifetime ($\tau$) is calculated using phenomenological theory [36,37]

$$\tau = \left[ (\alpha \omega + \gamma \mu_0 H_\Delta) \frac{\partial \omega}{\partial \omega_H} \right]^{-1}. \quad (10)$$

Now, with the knowledge of spin-wave eigenstates and their lifetimes, we can proceed to the calculation of the Bloch function. A complex circular dynamic magnetization ($\mathcal{M}$) is assumed (the possible ellipticity of the dynamic magnetization will be assumed later by calculating the amplitude of the spin wave mode).

$$\mathcal{M}(\omega, \boldsymbol{Q}) = M'_x(\omega, \boldsymbol{Q}) + i M'_y(\omega, \boldsymbol{Q}). \quad (11)$$

This complex circular magnetization now describes the possible spin-wave resonances. If we assume only one spin wave mode with a specific in-plane wavevector, the frequency-dependent complex magnetization will have Lorentzian shape [38],

$$\mathcal{M}(\omega, \boldsymbol{Q}) \propto \frac{1}{(\omega_0 - \omega)^2 + \left(\frac{2}{\tau}\right)^2} + i \frac{1}{(\omega_0 - \omega)^2 + \left(\frac{2}{\tau}\right)^2}, \quad (12)$$

where the width of the resonance is given by the spin-wave lifetime $\tau$. To obtain the Bloch function, we take the absolute value out of the complex magnetization and, as we are interested in thermally excited spin waves, correct the resulting function for the Bose-Einstein distribution (since we assume that all Lorentzian oscillators have the same amplitude)

$$\mathcal{D}(\omega, \boldsymbol{Q}) \propto \sqrt{2 n_{BE}(\omega)} \frac{1}{(\omega_0 - \omega)^2 + \left(\frac{2}{\tau}\right)^2}, \quad (13)$$

where the Bose-Einstein distribution is given by

$$n_{BE} = \frac{1}{\exp\left(\frac{\hbar \omega - \mu}{k_b T}\right) - 1}, \quad (14)$$

$\mu$ is a chemical potential, $k_b$ is the Boltzman constant, $\hbar$ is the reduced Planck constant, and $T$ is a thermodynamic temperature.

To obtain the out-of-plane and in-plane magnetization [39], we need to multiply the solution by the spin-wave profile amplitude

$$M_{IP}(\boldsymbol{Q}, \omega, \xi) = m_{Q,\rho}(\boldsymbol{Q}, \omega, \xi) \mathcal{D}(\boldsymbol{Q}, \omega), \quad (15a)$$

$$M_{OOP}(\boldsymbol{Q}, \omega, \xi) = i m_{Q,\xi}(\boldsymbol{Q}, \omega, \xi) \mathcal{D}(\boldsymbol{Q}, \omega). \quad (15b)$$

In Fig. 4(a) we show the dispersion relation for all directions in the in-plane magnetized 30 nm-thick Permalloy layer in 10 mT external field applied along the $y$ axis. In Figs. 4(b)–4(e) Bloch functions are shown for selected frequencies. At 3 GHz [Fig. 4(b)] the resonance is only in the so-called backward volume direction (parallel to the applied field), i.e., there is no intensity on the cross section through $k_y$. For higher frequencies (above ferromagnetic resonance frequency) the resonance for DE geometry appears, see Figs. 4(c)–4(e). With increasing frequency, one can observe that the linewidth in $k_x$, $k_y$ space becomes wider. This is especially visible for the spin-wave modes with low wave numbers, see Figs. 4(b)–4(d).

## V. CALCULATION OF THE POLARIZATION

In the continuum model, the inelastic scattering (shift of the frequency of the scattered light) is caused by the dynamic (time-dependent) susceptibility. In the studied case, we assume magneto-optical coupling which is usually described by two contributions: the so-called Voigt effect, which is linear in magnetization, and the Cotton-Mouton effect, which is quadratic in magnetization. The resulting susceptibility ($\hat{\chi}_{MO}$) can be written as

$$\hat{\chi}_{MO} = \begin{pmatrix} 0 & iM_z Q & -iM_y Q \\ -iM_z Q & 0 & iM_x Q \\ iM_y Q & -iM_x Q & 0 \end{pmatrix} + \begin{pmatrix} B_1 M_x^2 & B_2 M_x M_y & B_2 M_x M_z \\ B_2 M_x M_y & B_1 M_y^2 & B_2 M_y M_z \\ B_2 M_x M_z & B_2 M_y M_z & B_1 M_z^2 \end{pmatrix}, \quad (16)$$

where $M_x, M_y, M_z$ denote the magnetization vector components, $B_1, B_2$ are Cotton-Mouton magneto-optical constants, and $Q$ is Voigt magneto-optical constant. For simplicity, in the presented analysis we assume $B_1 = B_2 = 0$. Please note, that in the cases where $B_1 \neq 0$ or $B_2 \neq 0$, the multiplication of two dynamic magnetization components results in the inelastic shift on the double of the magnon frequency. This process can be described by the scattering of one photon on two magnons. But this effect is very weak in ferromagnets, and the measured signal would probably be too low to be observed in experiments [40].

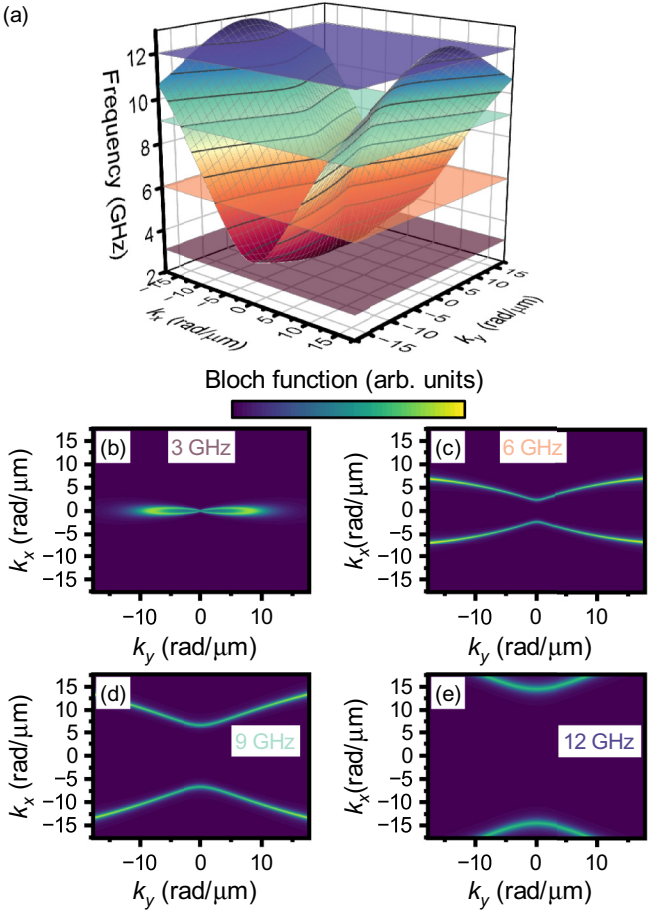

FIG. 4. Full in-plane dispersion and Bloch functions of spin waves. (a) Spin wave dispersion relation for all propagation directions for in-plane magnetized 30 nm-thick Permalloy film. The transparent planes depicts the frequency positions of the calculated Bloch functions. (b)–(e) Calculated Bloch functions for 3 GHz (b), 6 GHz (c), 9 GHz (d), and 12 GHz (e). The external field was set to 10 mT and $\mu = -6.6 \cdot 10^{-25} = \frac{-1\text{THz}}{h}$.

In this model, the incident electric field $\boldsymbol{E}_\text{d}$ probes the dynamic modulation of the susceptibility via magneto-optical coupling, which gives the polarization inside the magnetic material in the form [5,41,42]

$$\boldsymbol{P}(t, \boldsymbol{r}) = \boldsymbol{E}_\text{d}(t, \boldsymbol{r})\hat{\chi}(t, \boldsymbol{r}), \quad (17)$$

where $\hat{\chi} = \hat{\chi}_\text{mat} + \hat{\chi}_\text{SW}(t, \boldsymbol{r})$ is a sum of the static material susceptibility $\hat{\chi}_\text{mat}$ and of the additional dynamic contribution caused by spin waves $\hat{\chi}_\text{SW}(t, \boldsymbol{r})$. Here we emphasize that modulations in the susceptibility caused by magnons are on a vastly different time scale than the optical cycle of the probing photons. Thus, from the light's point of view, the situation is similar to scattering on a static grating. As a consequence, there is a mixing of the frequencies in both, temporal and spatial domains, namely

$$\boldsymbol{P}(\omega, \boldsymbol{k}_\text{p}, z) = \hat{\chi}(\omega_\text{m}, \boldsymbol{k}_\text{m}, z)\boldsymbol{E}_\text{d}(\omega - \omega_\text{m}, \boldsymbol{k}_\text{p} - \boldsymbol{k}_\text{m}, z), \quad (18)$$

where $\omega$ is the frequency of the induced polarization, $\boldsymbol{k}_\text{p}$ is its in-plane (parallel to the magnetic layer) wave vector, while $\omega_\text{m}$ and $\boldsymbol{k}_\text{m}$ are their magnon counterparts. This equation represents the convolution of the Fourier images of the susceptibility $\hat{\chi}$ and the driving field $\boldsymbol{E}_\text{d}$. The vertical profile of the dynamic magnetization (along $z$) depends on the exact geometry and mode of the spin wave and should be taken into account for precise calculation. In the analysis presented here, we ignore this dependence. However, this approximation may be insufficient in the case where there is a strong dependence of the mode profile on the $z$ coordinate, comparable to the penetration depth of the light. This can be particularly true for transparent materials such as ytrium-iron-garnet (YIG), where the penetration length can be greater than the sample thickness. However, this dependence can be introduced by dividing the magnetic layer into several thin sublayers and treating each as an independent polarization source. Another approximation that can be safely made, since $\omega \gg \omega_\text{m}$, is to drop the exact dependence of the driving field on $\omega_\text{m}$. As a consequence, it is sufficient to calculate the driving field at a single frequency $\omega$, which can significantly reduce the computation time of the presented model, especially when the driving field requires numerical simulation.

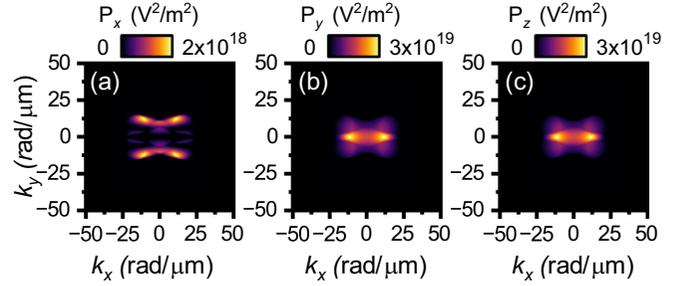

FIG. 5. The calculation of the induced polarization at 6 GHz in 10 mT. (a)–(c) The squared Fourier-transformed induced polarization $x$- (a), $y$- (b), and $z$-component (c).

We have calculated the resulting polarization considering only the linear Voigt contribution using Eq. (18). We can observe that the polarization current in the $x$ component [Fig. 5(a)] has an order of magnitude lower intensity compared to the $y$ and $z$ components [Figs. 5(b) and 5(c)]. Note, that polarization currents are formed with wave vectors inaccessible to free light (in this case $k > 10\,\text{rad}/\mu\text{m}$).

## VI. TRANSITION TO THE FAR FIELD AND CALCULATION OF THE BLS SIGNAL

The polarization vector in Eq. (18) acts as a local radiation source that eventually generates the detected BLS signal. The contribution of a given spatial frequency to this signal is determined by its ability to efficiently couple into the free space continuum and pass through the optical setup towards the detector. In the case of so-called $k$-resolved BLS, the spatial frequencies are given by the wave vector of the incident light and its angle with respect to the sample normal. In the case of micro-focused BLS, the range of spatial frequencies that can reach the detector is mainly limited by the numerical aperture of the used objective lens.

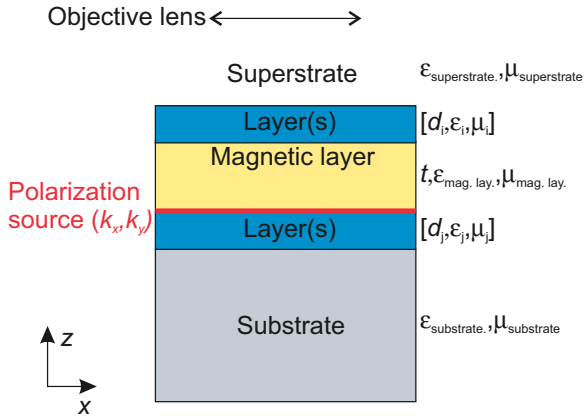

FIG. 6. Schematics of the geometry for far-field transition. The light is focused on the magnetic layer by the objective lens, which is in the superstrate (usually air). We can assume an arbitrary number of layers above (coating) and beneath the magnetic layer. The polarization (radiation source) is placed in the middle of the magnetic layer. For simplicity, we assume for all materials $\mu = 1$.

The transition from the polarization source to the far field can be expressed mathematically using the Green's function formalism

$$\mathbf{E}_{\mathrm{FF}}(\omega, \mathbf{k}_{\mathrm{p}}) = \hat{\mathbf{G}}(\omega, \mathbf{k}_{\mathrm{p}}, \mathbf{k}'_{\mathrm{p}}) \mathbf{P}(\omega, \mathbf{k}'_{\mathrm{p}}). \quad (19)$$

The dyadic Green's function $\hat{\mathbf{G}}(\omega, \mathbf{k}_{\mathrm{p}}, \mathbf{k}'_{\mathrm{p}})$ embodies the response of a system to a local source and can fully account for the presence of any scattering object, substrate effects or complex geometries. It relates the spatial frequencies $(\omega, \mathbf{k}'_{\mathrm{p}})$ of the polarization vector to the spatial frequencies $(\omega, \mathbf{k}_{\mathrm{p}})$ of the generated electric field vector and the term dyadic highlights its second rank tensor nature.

Here we provide a representation of the dyadic Green's function for a (possibly) multilayer continuous film (stratified medium) with Gaussian illumination through the objective lens. We use the theory developed by Sommerfield and Weyl [30–32].

We assume infinite layers in the $xy$ plane with broken symmetry in the $z$ direction, as shown in Fig. 6. The superstrate (air in our calculation) and the substrate are also semi-infinite in the $z$ direction. Between them we assume a stack of layers, at least one of which is magnetic.

The dyadic Green's function providing the electric field at the position of the polarization source and expressed in terms of individual plane waves reads

$$\hat{\mathbf{G}}(\mathbf{k}_{\mathrm{p}}) = \frac{\mathrm{i}\mu_0 \omega^2}{2} \iint_{-\infty}^{\infty} \mathrm{d}^2 k'_p \hat{\mathbf{M}}^{\pm}, \quad (20)$$

where

$$\hat{\mathbf{M}}^{\pm} = \frac{1}{k_s^2 k_{\mathrm{zs}}} \begin{pmatrix} k_s^2 - k_x^2 & -k_x k_y & \pm k_x k_{\mathrm{zs}} \\ -k_x k_y & k_s^2 - k_y^2 & \pm k_y k_{\mathrm{zs}} \\ \pm k_x k_{\mathrm{zs}} & \pm k_y k_{\mathrm{zs}} & k_s^2 - k_z^2 \end{pmatrix}, \quad (21)$$

$k_s$ is a wave vector in the magnetic material, and $k_{\mathrm{zs}} = \sqrt{k_s^2 - k_x^2 - k_y^2}$ is its longitudinal projection. The choice of the sign in (21) depends on whether we are dealing with forward-propagating (+) or backward-propagating (−) waves (we should stress that the polarization source radiates into both half-spaces). This dyadic Green's function allows us to calculate the amplitudes of both propagating and evanescent waves from all orientations of the polarization **P**. To properly account for all the effects of reflection and refraction at the interfaces between the different layers, each plane wave must be multiplied by an appropriate Fresnel coefficient. In the case of a single interface between two media indexed 1 and 2, the Fresnel reflection $[r(k_x, k_y)]$ and transmission $[t(k_x, k_y)]$ coefficients for $p$ and $s$ polarized waves are given by the following set of equations:

$$r_{\mathrm{s}}(k_x, k_y) = \frac{k_{z1} - k_{z2}}{k_{z1} + k_{z2}} \quad (22\mathrm{a})$$

$$t_{\mathrm{s}}(k_x, k_y) = \frac{2k_{z1}}{k_{z1} + k_{z2}} \quad (22\mathrm{b})$$

$$r_{\mathrm{p}}(k_x, k_y) = \frac{\epsilon_2 k_{z1} - \epsilon_1 k_{z2}}{\epsilon_2 k_{z1} + \epsilon_1 k_{z2}} \quad (22\mathrm{c})$$

$$t_{\mathrm{p}}(k_x, k_y) = \frac{2\epsilon_2 k_{z1}}{\epsilon_2 k_{z1} + \epsilon_1 k_{z2}} \sqrt{\frac{\epsilon_1}{\epsilon_2}} \quad (22\mathrm{d})$$

By employing the *transfer-matrix method*, the above coefficients can be generalized to provide the amplitudes of the waves at any position within the stack, including the substrate and the superstrate [43,44].

Since the collection of the BLS signal takes place in the far-field and is limited only to the propagating part of the plane wave spectrum, it is useful to recast the tensor (21) into a form that better reflects the different treatment of the $p$- and $s$-polarized wave: changing the output basis (i.e., the rows) of the dyadic Green's function to spherical coordinates and recalling that the radial component of the propagating waves within an isotropic medium vanishes, we can break (21) into two distinct contributions:

$$\hat{\mathbf{M}}_{\mathrm{s}}^{\pm} = \frac{1}{k_{\mathrm{zs}}\sqrt{k_x^2 + k_y^2}} \begin{pmatrix} -k_y & k_x & 0 \\ 0 & 0 & 0 \end{pmatrix} \quad (23\mathrm{a})$$

$$\hat{\mathbf{M}}_{\mathrm{p}}^{\pm} = \begin{pmatrix} 0 & 0 & 0 \\ \pm \frac{k_x}{k_s \sqrt{k_x^2 + k_y^2}} & \pm \frac{k_y}{k_s \sqrt{k_x^2 + k_y^2}} & -\frac{\sqrt{k_x^2 + k_y^2}}{k_s k_{\mathrm{zs}}} \end{pmatrix}. \quad (23\mathrm{b})$$

Note that the new output basis ensures that the resulting electric field vector now has only two components, $E_s$ and $E_p$, which are directly related to the amplitudes of the propagating $p$- and $s$-polarized waves generated within the magnetic layer.

By replacing (21) with (23) and introducing corresponding Fresnel coefficients, we finally obtain an expression for the dyadic Green's function that fully accounts for the reflections within our system and the subsequent out-coupling of the emitted radiation to the far-field

$$\hat{\mathbf{G}}(\mathbf{k}_{\mathrm{p}}) = \frac{\mathrm{i}\mu_0 \omega^2}{2} \iint_{-\infty}^{\infty} \mathrm{d}^2 k'_{\mathrm{p}}$$
$$\times [t_{\mathrm{s}}^+ \hat{\mathbf{M}}_{\mathrm{s}}^+ + t_{\mathrm{s}}^- \hat{\mathbf{M}}_{\mathrm{s}}^- + t_{\mathrm{p}}^+ \hat{\mathbf{M}}_{\mathrm{p}}^+ + t_{\mathrm{p}}^- \hat{\mathbf{M}}_{\mathrm{p}}^-]. \quad (24)$$

After inserting (18) into (19) and recalling the fact that in the absence of a scattering object, there is no momentum transfer as the light leaves the stratified medium, the dark-field angular spectrum becomes

$$\mathbf{E}_{\mathrm{FF}}(\omega_{\mathrm{m}}\mathbf{k}_{\mathrm{m}}, \omega, \mathbf{k}_{\mathrm{p}}) = \int d^2\,\hat{\mathbf{G}}(\omega, \mathbf{k}_{\mathrm{p}})\hat{\chi}(\omega_{\mathrm{m}}, \mathbf{k}_{\mathrm{m}})\mathbf{E}_{\mathrm{d}}(\omega, \mathbf{k}_{\mathrm{m}}). \quad (25)$$

Another important aspect of the BLS detection process is the limited area from which the signal is collected. This is equivalent to the statement that only rays virtually parallel to the microscope's optical axis can successfully reach the detector. Assuming that the collection spot has a Gaussian spatial profile $h(x, y) = e^{-(x^2+y^2)/w_c^2}$, where $w_c$ is a waist of the detection spot. The detectable portion of the far-field radiation amounts to

$$\mathbf{E}_{\mathrm{FF}}(\mathbf{r}_\parallel) = h(\mathbf{r}_\parallel)\int_{k_{\mathrm{p}}\leqslant k_0\mathrm{NA}} d^2 k_{\mathrm{p}}\, e^{i\mathbf{k}_{\mathrm{p}}\cdot\mathbf{r}_\parallel}\mathbf{E}_{\mathrm{FF}}(\mathbf{k}_{\mathrm{p}}), \quad (26)$$

where the integration limits reflect the restrictions placed on the spatial frequencies by the numerical aperture of the objective lens.

Finally, to estimate the strength of the BLS signal at a particular frequency $\omega_{\mathrm{m}}$, one has to sum up the contributions from all magnons (i.e., integrate over $\mathbf{k}_{\mathrm{m}}$). The exact nature of this summation depends on the coherence properties of the magnon population. In the case of thermal magnons (which are inherently incoherent), the proper procedure is to add intensities originating from individual magnon contributions. In the case of coherent magnons (excitation either by microstrip antenna or, e.g., by vortex core motion,), one must account appropriately for the phase and sum of all waves before calculating intensities. The modeled BLS signal for thermal magnons (integration after squaring) reads

$$\sigma(\omega_{\mathrm{m}}) = \int d^2 r_\parallel \int d^2 k_{\mathrm{m}} \left| h(\mathbf{r}_\parallel) \int_{k_{\mathrm{p}}\leqslant k_0\mathrm{NA}} d^2 k_{\mathrm{p}}\, e^{i\mathbf{k}_{\mathrm{p}}\cdot\mathbf{r}_\parallel} \int d^2 k'_{\mathrm{p}} \right.$$
$$\left. \times \hat{\mathbf{G}}(\mathbf{k}_{\mathrm{p}})\,\hat{\chi}(\omega_{\mathrm{m}}, \mathbf{k}_{\mathrm{m}})\mathbf{E}_{\mathrm{d}}(\mathbf{k}_{\mathrm{m}}) \right|^2, \quad (27)$$

while for coherent magnons (integration before squaring), we get

$$\sigma(\omega_{\mathrm{m}}) = \left| \int d^2 r_\parallel \int d^2 k_{\mathrm{m}}\, h(\mathbf{r}_\parallel) \int_{k_{\mathrm{p}}\leqslant k_0\mathrm{NA}} d^2 k_{\mathrm{p}}\, e^{i\mathbf{k}_{\mathrm{p}}\cdot\mathbf{r}_\parallel} \int d^2 k'_{\mathrm{p}} \right.$$
$$\left. \times \hat{\mathbf{G}}(\mathbf{k}_{\mathrm{p}}, \mathbf{k}'_{\mathrm{p}})\,\hat{\chi}(\omega_{\mathrm{m}}, \mathbf{k}_{\mathrm{m}})\mathbf{E}_{\mathrm{d}}(\mathbf{k}'_{\mathrm{p}} - \mathbf{k}_{\mathrm{m}}) \right|^2. \quad (28)$$

## VII. COMPARISON BETWEEN THE MODEL AND EXPERIMENTS

In order to validate the presented theory, we compared the predictions of the model with experimentally obtained data. First, we compared the modeled and experimentally obtained BLS spectra in different external magnetic fields. In the second subsection we compared the experimentally estimated detection sensitivity with the model predictions.

### A. Thermal $\mu$-BLS

By performing $\mu$-BLS measurements of the thermally excited spin waves, we check the agreement of Eq. (27) with experimental data. All measurements were performed on a 31.5 nm-thick NiFe layer. For modeling we used the same set of material parameters as presented in [20]. We show the experimental data in Fig. 7(a) and the modeled spectra in Fig. 7(b). The amplitude of the modeled signal has been adjusted to match the experimental data. The model and experiment are in very good agreement. The experimental and modeled spectra show a fundamental spin wave mode (lower frequency) and a first perpendicular standing spin wave (PSSW, higher frequency) mode. In both experiment and model we observe a decrease in the amplitude of both modes with increasing magnetic field and a frequency narrowing of the fundamental mode spectra.

Figures 7(c) and 7(d) show the modeled and measured spectra of the fundamental mode at 10 mT and 558 mT. In the lower panel the peak has a sharp rising edge. This is caused by the shallow dispersion relation of the backward volume dispersion branch, see the light blue dashed line in Fig. 7(c). This shallow dispersion has a low group velocity and thus a high density of states around the ferromagnetic resonance (FMR) frequency. This results in a distinct peak in the BLS spectra at this frequency. In contrast, at high magnetic fields (such as 558 mT) the situation is much more symmetrical around the FMR frequency. There is not much difference in the group velocity (and density of states) for spin waves propagating in the Damon-Eshbach and backward volume directions. For this reason, the BLS spectrum resembles a near-Gaussian shape with a maximum located at the point of the FMR frequency. Please note, that in the studied case, the maximum signal in the peak originating from fundamental mode always lies in the position of FMR frequency. The amplitude of the signal originating from perpendicular standing spin waves was adjusted to match the experimental data.

### B. Coherent $\mu$-BLS

To test the coherently excited spin wave model, we fabricated a 180 nm wide microstrip antenna on the NiFe layer. We measured the intensity of the BLS while sweeping the excitation frequency with a set power of 10 dBm in the external magnetic field of 50 mT, which was directed parallel to the excitation microstrip, so that the spin waves propagate in the Damon-Eshbach geometry. The spectra were acquired at a distance of 1 μm from the antenna. The measured data are shown in Fig. 8(a).

In a simplified view of this one-dimensional geometry (where the spin waves are only excited by the excitation antenna with a single wave vector and frequency), the acquired BLS signal can be expressed as

$$\sigma(\mathbf{k}) = \Gamma_{\mathrm{exc}}(\mathbf{k})\Gamma_{\mathrm{det}}(\mathbf{k}), \quad (29)$$

where $\Gamma_{\mathrm{exc}}(\mathbf{k})$ is the excitation efficiency given by the antenna geometry and $\Gamma_{\mathrm{det}}(\mathbf{k})$ is a detection efficiency given by the optical setup. The excitation efficiency can be roughly estimated as the Fourier transform of the in-plane magnetic field generated by the microstrip antenna. For our geometry this is shown as the red line in Fig. 8(a).

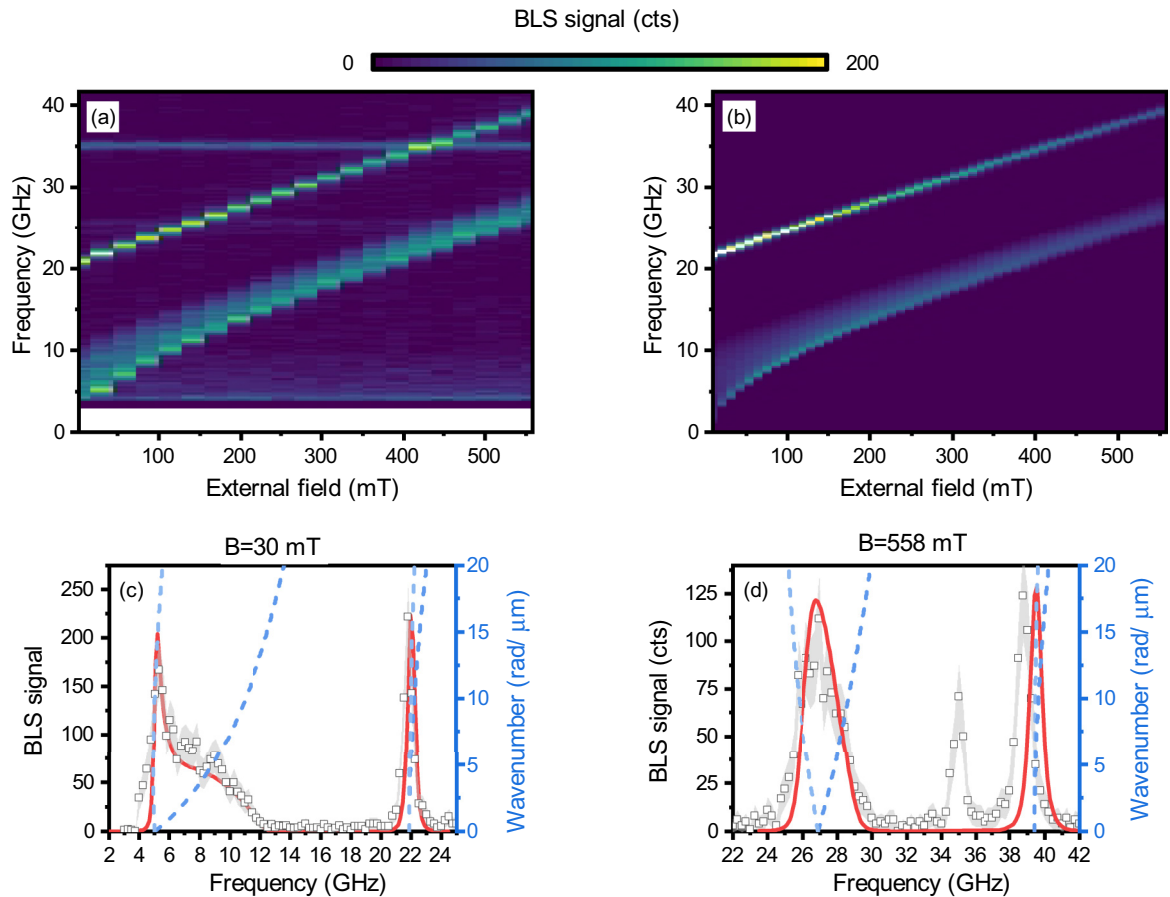

FIG. 7. Comparison between the experimentally obtained data and theoretical modeling of thermally excited spin waves. All data are for 31.5 nm-thick NiFe layer. (a), (b) BLS spectra in dependency to external magnetic field. Panel (a) shows the experimentally obtained data, and panel (b) shows the data modeled with Eq. (27). (c), (d) Experimentally obtained and modeled BLS spectra for the external field of 30 mT (c) and 558 mT (d). The signal visible in panel (a) and (d) at 34 GHz does not have any dependency on the applied magnetic field, and thus, it does not originate from spin waves.

Now, knowing the experimentally obtained BLS intensity $\sigma(\mathbf{k})$ and the calculated $\Gamma_{exc}(\mathbf{k})$, we estimated the detection function $\Gamma_{det}(\mathbf{k})$ of our $\mu$-BLS, see blue squares in Fig. 8(b). This sensitivity can also be modeled using Eq. (28), see black solid line in Fig. 8(b). The modeled and experimentally obtained detection functions are in good agreement, and both have a detection edge (1 % of detection sensitivity) at $\approx 12\,\mathrm{rad}^{-1}\mu\mathrm{m}$.

## VIII. APPLICATIONS OF THE MODEL

In this section we discuss four examples of applications of the presented model. All four cases are calculated for a continuous thin film (single or double layer) and can therefore be fully solved by the semi-analytical method presented here. In the first example, we calculate the detection sensitivity for coherently excited spin waves as a function of the numerical aperture of the objective lens. In the second example, we calculate the thermal signal and its total intensity as a function of sample thickness and show that there is an optimal thickness of the magnetic thin film for which we can obtain the maximum BLS signal. In the third example, we demonstrate the possibility of increasing (and decreasing) the BLS signal by appropriate selection of a silicon dioxide cover layer, and in the fourth example, we calculate the decrease of the thermal signal as a function of the thickness of a platinum cover layer.

### A. Detection sensitivity to coherent spin waves in dependence to the numeric aperture of the objective lens

The numeric aperture has a pronounced effect on the formed polarization in the material through the change of the driving field $\mathbf{E_d}$ [see Eq. (17)], but it also determines the range of wavevectors that can reach the detector [see Eq. (26)]. For the detection sensitivity in $\mu$-BLS experiments, numeric aperture is crucial. The higher the numeric aperture is, the higher wave numbers of spin waves can be detected. The calculated $\mu$-BLS spectra of 30 nm-thick NiFe layer with a varying numeric aperture of the objective lens and filling factor 2 [45] are shown in Fig. 9(a), and selected spectra are separately plotted in Fig. 9(b). Indeed, we can observe the expected trend of broadening detection sensitivity towards the higher wave numbers [18]. We extracted the value, where the detection sensitivity dropped to the 10 % of the maximal value, see Fig. 9(c). The observed behavior is linear. The simple analytic formula can be found to approximate this detection *threshold*. We start with an approximate formula for waist of the

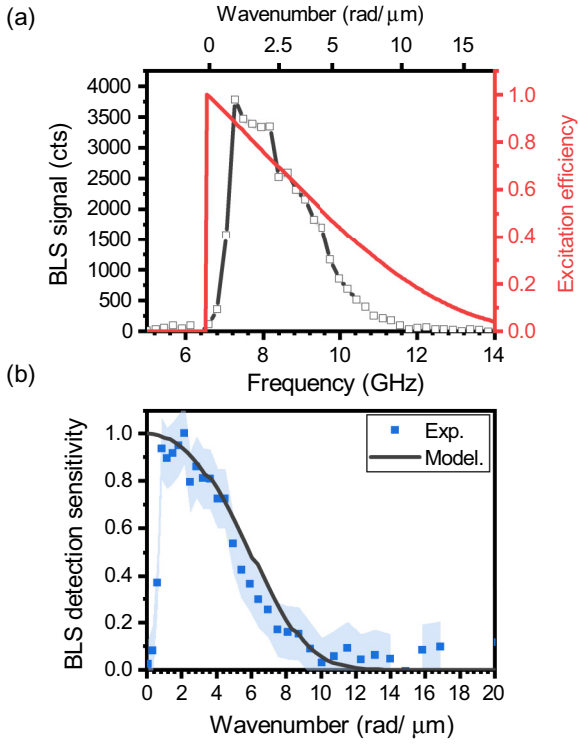

FIG. 8. Comparison between the experimentally obtained data and theoretical modeling of coherently excited spin waves. (a) Experimentally measured BLS intensity (black line) of coherently excited spin waves 1 μm away from 180 nm-wide antenna. The red line shows the calculated excitation efficiency of the used antenna. The top x-axis shows the wave number calculated by using the dispersion relation. (b) Squares shows extracted detection efficiency of μ-BLS. The solid line shows calculation based on Eq. (28).

beam spot

$$2w_0 = \frac{2\lambda}{\pi \mathrm{NA}}, \quad (30)$$

where $w_0$ is radius of a beam spot (waist), $\lambda$ is a wavelength, and NA is a numeric aperture of objective lens. Now, by taking the Fourier transform of the beam spot, we can estimate the detection sensitivity threshold as a half-width-at-the-tenth-of-maximum (HWTM) of the transformed expression. This leads to

$$\mathrm{HWTM} = \sqrt{2\ln(2)}\frac{\pi}{\lambda}\mathrm{NA} \approx 6.72 \frac{\mathrm{NA}}{\lambda}. \quad (31)$$

This simple analytic formula agrees reasonably well with the full theoretical description and can be used for quick estimations of detectable wave numbers.

### B. Thermal signal in dependence to the thickness of NiFe layer

The change in the thickness of the film does not affect the frequency of ferromagnetic resonance ($k = 0$). However, it dramatically affects spin wave dispersion, particularly spin wave group velocities. Also, the spin-wave signal is proportional to the interaction volume, which is decreased when the thickness of the film is lowered. To obtain indecent electric field, the light propagation in the magnetic medium is assumed to be exponentially attenuated on the way to the material. This approach is a rather crude approximation as the same attenuation is assumed for all directions of wavevectors, and no influence on the shape of the electric field is considered. This approximation is introduced as a factor ($\mathfrak{F}$),

$$\mathfrak{F} = \int_0^{t_{\mathrm{mag}}} \exp{(-k_{\mathrm{exc}}k_0 z)^2} \mathrm{d}z, \quad (32)$$

where $t_{\mathrm{mag}}$ is thickness of the magnetic layer, $k_{\mathrm{exc}}$ is the extinction coefficient of the magnetic material, and $k_0$ is free space wave number of the used light. The whole equation for the BLS signal then reads as

$$\sigma(\omega_{\mathrm{m}}) = \int \mathrm{d}^2 r_{\parallel} \int \mathrm{d}^2 k_{\mathrm{m}} \bigg| h(\mathbf{r}_{\parallel}) \int_{k_{\mathrm{p}} \leqslant k_0 \mathrm{NA}} \mathrm{d}^2 k_{\mathrm{p}}\, e^{i\mathbf{k}_{\mathrm{p}} \cdot \mathbf{r}_{\parallel}}$$
$$\times \int \mathrm{d}^2 k'_{\mathrm{p}} \hat{\mathbf{G}}(\mathbf{k}_{\mathrm{p}}, \mathbf{k}'_{\mathrm{p}})\, \hat{\boldsymbol{\chi}}(\omega_{\mathrm{m}}, \mathbf{k}_{\mathrm{m}}) \mathfrak{F} \mathbf{E}_{\mathrm{d}}(\mathbf{k}'_{\mathrm{p}} - \mathbf{k}_{\mathrm{m}}) \bigg|^2. \quad (33)$$

This equation approximates magnetization dynamics with a uniform precession angle across the thickness of the magnetic layer. This is quite a strong approximation, especially in the case of the quantized thickness modes. But it also neglects the case of nonzero wave numbers or nonzero spin pinning on the layer boundaries.

The resulting BLS spectra for NiFe layers with varying thicknesses are shown in Fig. 9. The NiFe is metallic and thus has strong attenuation of the light in the material, with an extinction coefficient of $k_{\mathrm{exc}} = 3.842$.

In low magnetic fields ($B_{\mathrm{ext}} = 10$ mT), the *turning point* in BV geometry is accessible by the $\mu$−BLS, and this exhibits itself as a strong signal in the lowest detected frequency as can be seen across all thicknesses in Fig. 9(d). With increasing thickness, we can observe broadening to higher frequencies. This is caused by the increase in the group velocity of the spin waves. In the high magnetic field ($B_{\mathrm{ext}} = 500$ mT), the BLS spectrum is much more symmetrical, see Fig. 9(e).

In both magnetic fields, when the thickness of the layer increases, the frequency of the first perpendicular-standing-spin-wave mode decreases and slowly reaches the value of ferromagnetic resonance. The higher perpendicular-standing-spin-waves were not considered.

In Fig. 9(f), the intensities of all four peaks are integrated. In integrated intensities of fundamental spin-waves (solid lines), we can see that the signal strength reaches the maximum the thickness of the magnetic layer around 25 nm for a field of 500 mT, and 50 nm for a field of 10 mT. This position of maximum signal is determined by the interplay between the contribution of larger interaction volume and from lower group velocity in thinner NiFe layers. However, the interaction volume does not increase linearly with the increasing thickness, and after reaching ≈70 nm is not increased at all due to the substantial decay of the light in the NiFe layer.

In the case of the first thickness mode (dashed lines), the intensity is almost zero for low thicknesses as the frequencies of these modes reach several terahertz. Due to the Bose-Einstein distribution, the population of the magnons on these frequencies is very low in comparison to the frequencies in the order of several gigahertz. In contrast, as the group velocity is increased in the range of the accessible wavevectors for thicknesses above 40 nm, this results in the decrease of the signal. These two factors cause the formation of the optimal thickness for the signal strength at ≈50 nm.

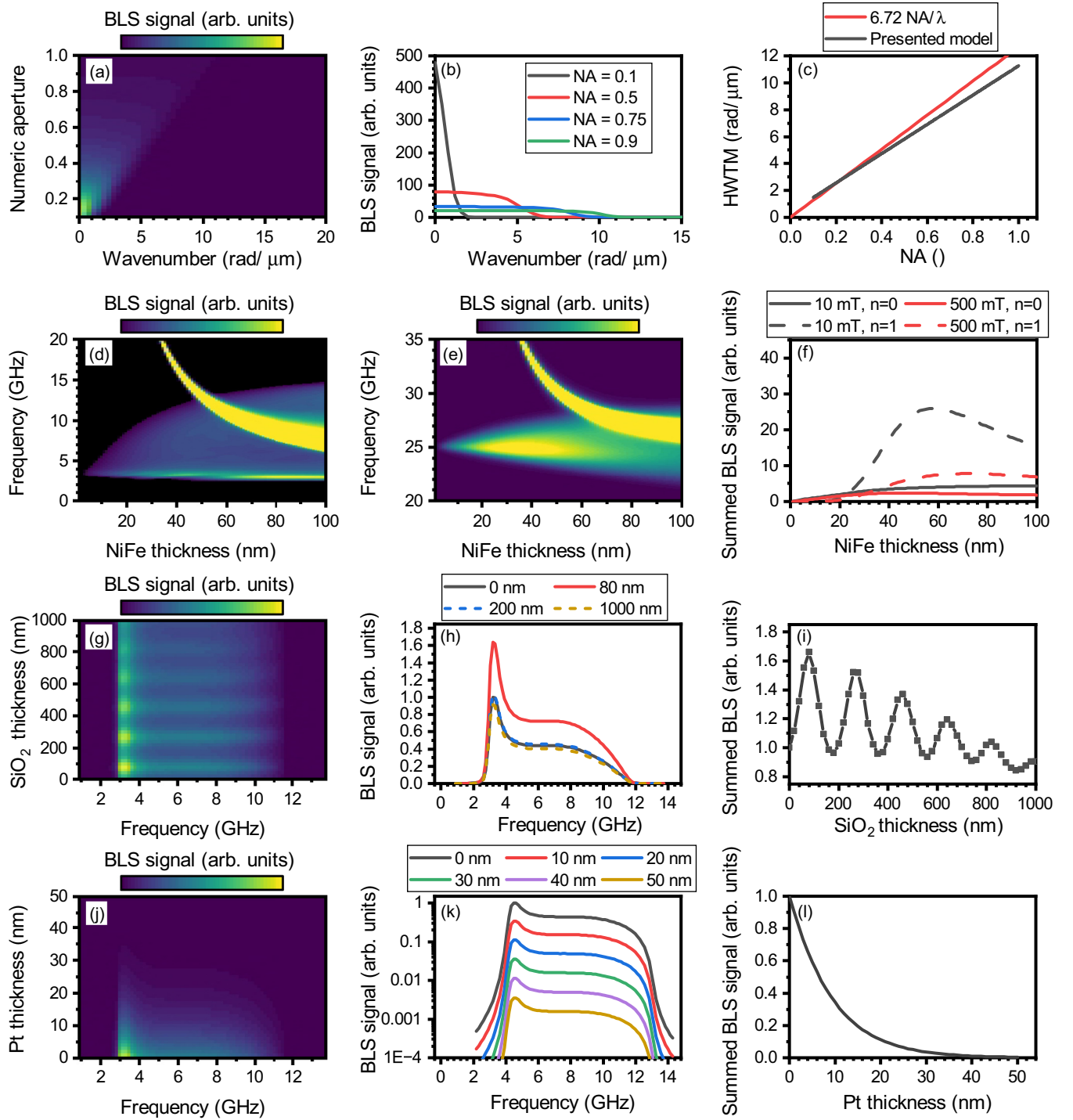

FIG. 9. Examples of applications of presented semianalytical model. (a)–(c) Dependence of the sensitivity of $\mu$-BLS to coherently excited spin waves. The panel (a) shows the BLS signal in dependence on the numeric aperture of the objective lens of coherently excited spin waves. The panel (b) shows individual BLS spectra for NA = 0.1, 0.5, 0.75, 0.9. The panel (c) shows the extracted threshold wave number (90 % of detection sensitivity) and a simple analytic formula. (d)–(f) Calculated $\mu$-BLS signal for NiFe layer with varying thickness. In panel (d) [(e)] the external field is set to 10 mT [500 mT]. The panel (f) shows the summed intensity of individual modes. (g)–(l) Calculated $\mu$-BLS spectra for NiFe layer covered by SiO$_2$ (g)–(i), and platinum (Pt) (j)–(l) with varying thickness. In panels (g), (j), the $\mu$-BLS spectra are shown in linear scale, and selected thicknesses are shown in linear scale for SiO$_2$ (h) and for Pt in logarithmic scale (k). In panel (i), (l) the summed $\mu$-BLS intensity is shown.

### C. Effect of the transparent cover layer

In some experimental scenarios, it is beneficial to put various materials on top of the magnetic layer to, i.e., engineer magnetic properties of the investigated layer or to protect it from the environment. These layers then affect the obtained $\mu$-BLS signal. With the presented model, we can estimate its impact. The model takes into account all of the reflections on material boundaries during the scattered light's way out and different adsorptions of the defined layers.

First, we consider the cover layer of $SiO_2$. We assume complex dielectric function $\epsilon = 2.1516 + i0.0058434$ [46]. Due to the constructive interferences in the cover layer, the $\mu$-BLS signal can be enhanced in specific thicknesses [47], see Figs. 9(g) and 9(h). This enhancement is homogeneous across all magnon frequencies. In Fig. 9(j), we can observe the period of the enhancement, that is roughly $190 \pm 20$ nm, which is below the wavelength of the used light in the $SiO_2$ (247 nm).

### D. Effect of the metallic cover layer

In Figs. 9(j) and 9(k) the spectra with different thickness of Pt cover layer are shown. The shape of the spectra is not affected by the presence of a platinum layer (we do not take into account any effect of platinum on magnetic properties, just its effect on light propagation). However, the intensity of the signal is decreased. With 50 nm of platinum the signal is decreased by approx. 3 orders of magnitude. The summed BLS signal has an exponential dependency on the platinum thickness, see Fig. (9).

### IX. CONCLUSION

The spectra obtained by $\mu$-BLS have so far only been analyzed qualitatively, without any insight into the formation of the signal. This severely limits the information that can be extracted from these experiments. Usually, only the intensity or the position of the BLS peaks are analyzed, but not their exact shape. In addition, many researchers have applied models developed for $k$-resolved BLS, which works with large spot sizes, to $\mu$-BLS, where spot sizes are usually diffraction-limited. However, this assumption often leads to incorrect conclusions, such as an overestimation of the detection sensitivity of $\mu$-BLS.

Here we present a theoretical model for the calculation of the $\mu$-BLS signal. This model opens up completely new ways of analyzing the spectra, including the analysis of their exact shape. Its calculation is fast (less than a few seconds per spectrum) and can therefore be used to fit any optical or magnetic parameter of the system. The model can also be used in conjunction with micromagnetic simulations to correctly interpret more complicated situations, such as nonlinear effects like parametric pumping. By knowing the exact sensitivity to specific wave vectors and by modeling the exact shape of the acquired BLS spectra, a whole new set of phenomena can be discovered and studied.

All data and code used to generate the presented figures are available in the Zenodo repository [48].

### ACKNOWLEDGMENTS

This research was supported by the Project No. CZ.02.01.01/00/22_008/0004594 (TERAFIT) and by the Grant Agency of the Czech Republic, Project No. 23-04120L. CzechNanoLab Project No. LM2023051 is acknowledged for the financial support of the measurements and sample fabrication at CEITEC Nano Research Infrastructure.